\documentstyle[aps,multicol,pre,eqsecnum,epsfig]{revtex}
\begin{document}
\title{Bulk dynamics for interfacial growth models}
\author{ Crist\'obal L\'{o}pez$^{1}$, Pedro L. Garrido$^2$ and 
Francisco de los Santos$^3$}

\address{
$^1$Instituto Mediterr\'{a}neo de Estudios Avanzados,
CSIC-Universitat de les Illes Balears, E-07071 Palma de Mallorca,
Spain \\ 
$^2$Instituto Carlos I de F\'{\i}sica Te\'orica y Computacional,
Universidad de Granada, E-18071 Granada, Spain \\
$^3$
Centro de F\'{\i}sica da Mat\'eria Condensada da Universidade de Lisboa, 
Av. Prof. Gama Pinto 2, P-1649-003 Lisboa, Portugal
}
\date{\today}
\maketitle

\begin{abstract}
We study the influence of the  bulk dynamics of a growing cluster of
particles on the properties of its interface. 
First, we  define
a {\it general bulk growth model} by means of a continuum Master equation for
the evolution of the bulk density field. This general model
just considers arbitrary addition of particles (though it can
be easily generalized to consider substraction) with no 
other physical restriction.
The corresponding 
Langevin equation for this bulk density
field is derived where the influence
of the  bulk dynamics is explicitly shown.
Finally, when it is assumed a  well-defined interface for the
growing cluster, the Langevin equation for
the height field of this interface
for some particular bulk dynamics is written. In particular, we obtain
the celebrated Kardar-Parisi-Zhang (KPZ) equation.
A Monte Carlo simulation illustrates the theoretical results.
\end{abstract}
\begin{multicols}{2}
\narrowtext

\section{Introduction}

In the last fifteen years there has been a great interest
in the study of the growth of surfaces by
dynamic processes based in addition and substraction of particles
(see, for example, \cite{G,KV,BS,HZ}).
For instance, the understanding of the conditions under which a growing
surface shows a rough structure is nowadays of the greatest importance
in the production of thin films and/or pure cristals.
Surface 
growth is usually studied by using lattice
models in which simple stochastic rules intend to mimic
the relevant phenomena. 
Their extensive computer simulation have been of a major importance in
characterizing and understanding the different morphologies that occur
in real experiments. However, due to the intrinsic limitations of computers
capabilities, some very interesting aspects
are usually subject to
inconclusive analysis and data interpretation.
In particular, let us remark
the inherent difficulty in the study of the
surface long time behavior and its scaling properties. Nevertheless, 
for this particular
aspect, analytical models seem to give us the answers to the questions 
that the computers fail to clarify.
These are mainly based in postulating 
a Langevin equation for the height
of the interface measured from a reference substrate. 	
Such Langevin equations intend to mimic the system microscopic dynamics and
its collective effect at large scales in space and time.
A general choice has the structure
\begin{eqnarray}
\partial_t h_t({\bf x})&&=\nu_1{\nabla}^2h_t+\nu_2
\vert{\nabla}h_t\vert^2+\nu_3{\nabla}^2
{\nabla}^2h_t
\nonumber\\ &&+\ldots +\eta_t({\bf x})\quad ,
\label{general}
\end{eqnarray}
where $h_t({\bf x})$ is the height of the interface at time $t$ at
the substrate position ${\bf x}\in R^d$ and  
$\eta_t({\bf x})$ is a 
white noise term.
It is generally assumed a one-to-one correspondence between
various terms in (\ref{general})
and different physical processes
(for example see the discusion in \cite{LDS} concerning a model
for epitaxial growth, or a general method in 
\cite{MMTB} to propose an equation such as 
(\ref{general}) by
using the reparametrization invariance symmetry).
The details of the microscopic processes that are assumed to be 
irrelevant at this
scale of observation are taken into account through
the values of the coefficients $\nu_1$, $\nu_2$, $\ldots$ and the
properties of the noise term $\eta_t$. 
Once the Langevin equation (\ref{general}) is defined, one
may apply renormalization group procedures to obtain different
universal properties and scaling behaviors.
The success of this scheme is clearly represented by the definition and
analysis of the Kardar-Parisi-Zhang
equation (KPZ) \cite{KPZ} which has been a clear breakthrough in the
study of the space-time asymptotic behavior of growth models.

Quite often surface
growth is a consequence of {\it bulk} dynamic processes.
A good example of this is provided by the growth of bacterial 
colonies where bacteria multiply in a nutrient environment, 
the shape of the colony being the moving interface \cite{vicsek}.
In general, the dynamics of the particles before and after
its aggregation
to a substrate
may influence the system interfacial behavior. 
For instance,
it may appear shadowing effects as it happens in DLA
processes and in thin film
growth \cite{KBR}, or they may induce different scaling regimes
depending on the time interval studied as it is shown in some molecular
beam epitaxy models in which a system bulk dynamics is defined
\cite{DLGK,SK}.
However, interface models are usually expressed in terms of a height field,
$h_t({\bf x}) \ge 0$. In doing so, 
bulklike contributions are neglected since only interfacial degrees
of freedom are being considered.
Unfortunately, the mathematical hurdles to deriving the phenomenological 
dynamics of interfaces from
stochastic bulk microscopic models 
is formidable, and a comprehensive theoretical picture is 
still lacking although a significant body of rather rigorous work has 
been devoted to the subject \cite{spohn}.

It is well known that the 
macroscopic behavior of systems at nonequilibrium states
exhibits a strong dependence on the functional
structure of its microscopic dynamics (for instance,
in the so-called {\it two-temperature Ising model} \cite{GM}
one finds that the phase diagram changes radically depending on
the analytic form for the probability of a spin-flip).
Nevertheless, one may expect that this strong relation between
microscopic dynamics and 
macroscopic behavior should dissappear near a 
renormalization group (RG) fixed point or in the scaling
regime where {\it universality} seems to guarantee that the
microscopic details are irrelevant (at least, one knows that
this is true when studying dynamic properties of equilibrium systems 
near a (RG) fixed point
\cite{HH}).
However, some recent results on 
the critical behavior of a nonequilibrium driven diffusive system
show that the microscopic dynamics may play a relevant role 
in the determination of the system universality class \cite{GSM}.
The influence of the microsocopic dynamical details into the
critical and non-critical properties of a nonequilibrium model
implies, in our opinion, that any a priori construction of a Langevin
equation as (\ref{general}) may occasionally disregard important features.

In this paper we introduce a quite general class
of nonequilibrium bulk growth models for which the 
aforementioned problems can be addressed
(there are in the literature some efforts in this
direction \cite{GGG,KMTB,PC}).
We shall define
a stochastic bulk local dynamics 
expressed by an appropiated 
continuum Master equation in which, for simplicity, 
only addition of particles is considered.
From the Master equation and using a truncated Kramers-Moyal 
expansion, we shall derive
a Langevin equation for the bulk degrees of freedom in which there is 
an explicit dependence on 
the analytic form of the rates.
In order to study surface properties, in a subsequent section 
we shall derive, from this bulk equation, an expression for the 
interfacial height field dynamics.
The illustrative example we take is that of the KPZ equation and,
for consistency, in that particular case we check our results 
by means of a Monte Carlo simulation.
Some other examples are then briefly commented and our conclusions
are given in the final section.
 
\section{Growth models: A general description}

\begin{figure}
\epsfig{file=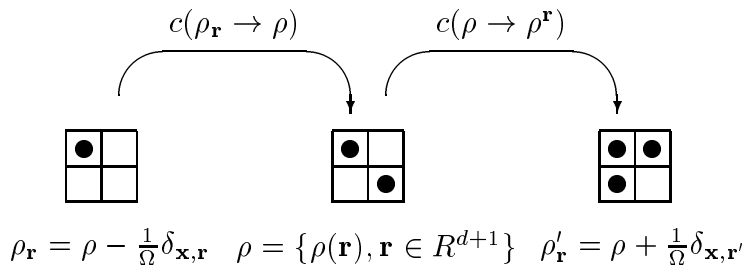,width=1.5\linewidth}\caption{Schematic 
representation of the proccess described by
the Master equation (\ref{MEGM}).}
\label{fig1}
\end{figure}

Let us consider a particle density field, $\rho({\bf x},\tau)$, ${\bf x}
\in R^{d+1}$, and assume that the probability distribution $P_\tau(\rho)$
associated with each field configuration obeys the following continuous 
Master equation
\begin{eqnarray}
\partial_{\tau} P_{\tau}({\bf \rho})=&&\int_{R^{d+1}}d{\bf r}\biggl[ 
c({\bf \rho}_{{\bf r}}\rightarrow
{\bf \rho})P_{\tau}({\bf \rho}_{{\bf r}})\nonumber\\
&& \qquad \qquad -
c({\bf \rho}\rightarrow{\bf \rho}^{{\bf r}})P_{\tau}({\bf \rho})
\biggr].
\label{MEGM}
\end{eqnarray}
Here, $c({\bf \rho}\rightarrow{\bf \rho}')$ is the probability per unit time 
(or transition rate) from one configuration ${\bf \rho}$ to 
another ${\bf \rho}'$, and $\rho_{{\bf r}}=\rho({\bf x})-\Omega^{-1}
\delta({\bf x} -{\bf r})$,
$\rho^{{\bf r}}=\rho({\bf x})+\Omega^{-1}\delta({\bf x}
-{\bf r})$. 
Note that the density field, $\rho$, can only grow in steps of
size $\Omega^{-1}$. This is consistent with a picture in which
$\rho({\bf r})$ is 
a particle density that results after coarse graining over 
blocks of size $\Omega$ centered around $\bf r$ in a lattice.
Therefore, the Master equation (\ref{MEGM}), so defined, could be thought of
as if it only 
described processes that add 
one particle per block of the lattice 
per elementary time step. This is schematically represented in figure 
\ref{fig1}.

Next, we choose the transition rates such that
\begin{equation}
c({\bf \rho}\rightarrow{\bf \rho}')
=w({\bf \rho};{\bf r})\label{dinform},
\end{equation}
namely, they are a function that
depends only on the initial configuration ${\bf \rho}$ and the specific
point where mass is added.
Now, let us assume that $\Omega$ is large enough (tipically,
it should be much bigger than any microcopic length scale present in
the original physical problem but much smaller than the 
correlation lenght of the system)
and expand the Master
equation (\ref{MEGM}) on invers powers of $\Omega$. Then, using the
expansions

\begin{equation}
c({\bf \rho}\rightarrow{\bf \rho}^{{\bf r}})=w({\bf \rho};
{\bf r}),\label{exp1}
\end{equation}
\begin{eqnarray}
c({\bf \rho}_{{\bf r}}\rightarrow{\bf \rho})&=&
w\big(\rho-\Omega^{-1}\delta(x-r);{\bf r}\big)\nonumber \\
&=&\sum_{m=0}^{\infty}
{{(-1)^m}\over{m!\Omega^m}}{{\delta^m}\over{\delta\rho({\bf r})^m}}
w({\bf \rho};{\bf r}),
\label{exp2}
\end{eqnarray} 
\begin{equation}
P_{\tau}({\bf \rho}_{{\bf r}})=\sum_{m=0}^{\infty}
{{(-1)^m}\over{m!\Omega^m}}{{\delta^m}\over{\delta\rho({\bf r})^m}}
P_{\tau}({\bf \rho}),
\label{exp3}
\end{equation} 
we get the so-called Kramers-Moyal expansion of the
Master equation (\ref{MEGM}) \cite{vkm},
\begin{equation}
\partial_{\tau}P_{\tau}({\bf \rho})=\int_{R^{d+1}}d{\bf r}\sum_{l=1}^{\infty}
{{(-1)^l}\over{l!\Omega^l}}{{\delta^l}\over{\delta\rho({\bf r})^l}}
\Big[w({\bf \rho};{\bf r})P_{\tau}({\bf \rho}) \Big].\label{KME}
\end{equation}

The next step we take is to keep only the first two terms in (\ref{KME}).
Then, we can write down the following Fokker-Planck equation
\begin{eqnarray}
\partial_{\tau}P_{\tau}({\bf \rho})=&&\int_{R^{d+1}}d{\bf r}\biggl[
-{{1}\over{\Omega}}{{\delta}\over{\delta\rho({\bf r})}}\nonumber \\
&& \qquad \qquad+
{1\over{2\Omega^2}}{{\delta^2}\over{\delta\rho({\bf r})^2}}
\biggr]\Big(w({\bf \rho};{\bf r})P_t({\bf \rho}) \Big).\label{FPE}
\end{eqnarray}
To control the goodness of such approximation we appeal to
{\it Kurtz theorem} \cite{KU}, by virtue of which 
when $\Omega\rightarrow\infty$, and for a given
time $T<\infty$ then
\begin{equation}
\sup_{\tau<T}\vert{\bf \rho}_{\tau}-{\bf \tilde{\rho}}_{\tau}
\vert\leq\zeta_\Omega^T{{\log\Omega}\over{\Omega}}.
\label{KT}
\end{equation}
where ${\bf \rho}_{\tau}$ and ${\bf \tilde{\rho}}_{\tau}$ 
are typical time trajectories on
phase space which are solutions of the exact Master equation (\ref{MEGM})
and the Fokker-Planck one (\ref{FPE}) respectively. $\zeta_\Omega^T$ is
a random variable whose distribution does not depend on $\Omega$ and
satisfying $\langle \exp(\lambda\zeta_\Omega^T)
\rangle<\infty$ for any constant $\lambda>0$.
That is, for a given fixed time $T$ one can always find a large enough
$\Omega$ such that the difference between solving exactly the master
equation or solving the truncated version of it, is of
order $\log\Omega/\Omega$ and therefore negligible.
Moreover, this bound is the best one and no new terms of the Kramers-Moyal
expansion give better results. However, when $T\rightarrow\infty$
one cannot control, in general, the accumulated influence of the 
neglected terms in the expansion. 
But, since the study of growth models is manily focused on the 
undestanding of their
evolution properties, due to Kurtz's theorem the Fokker-Planck
equation (\ref{FPE}) 
is a valid theoretical starting point.

Lastly, the Fokker-Planck equation (\ref{FPE}) is equivalent
in the Stratonovich sense to the Langevin equation \cite{vkm}
\begin{eqnarray}
\partial_t \rho({\bf r},t)=&&w({\bf \rho};{\bf r})-
{1\over{4\Omega}}{\delta\over{\delta\rho({\bf r})}}w({\bf \rho};{\bf r})
\nonumber\\+&&{1\over{\Omega^{1/2}}}w({\bf \rho};{\bf r})^{1/2}
\nu_t({\bf r})
\label{LE}
\end{eqnarray}
where $t=\Omega^{-1}\tau$,
$\nu_t({\bf r})$ is a white noise with zero mean and
$\langle\nu_t({\bf r})\nu_{t'}({\bf r}') \rangle=
\delta({\bf r}-{\bf r}')\delta(t-t')$.
We recall that $w({\bf \rho};{\bf r})$ is the
probability 
per unit time of adding a particle (of mass $\Omega^{-1}$) at point ${\bf r}$.
If we considered the posibility of particle substractions then
we should do the following substitutions in equation (\ref{LE}):
$w\to w_+-w_-$ in the first term and and $w\to w_++w_-$ in the last
two terms. 
$w_{+(-)}\equiv w({\bf \rho};{\bf r})_{+(-)}$ is the probability
per unit time
of adding (substracting) a particle.

Before we proceed to the
next section, let us remark that
 a) eq. (\ref{LE}) describes the evolution
of the bulk density field of a system that growths by addition of
particles with, in principle, no other physical  assumptions,
and b) the
Langevin equation (\ref{LE}) depends directly on
the functional form of the bulk rates.
The  election of these bulk rates then 
provides the physical 
restrictions for the particular 
growth model that is going to be specifically modelled.
Also, it is remarkable that the 
influence of the bulk dynamics  affects 
the noise term through a nontrivial factor.

\section{Height dynamics: KPZ equation and Monte Carlo simulation.}

We proceed to single out the interfacial degrees of freedom of equation
(\ref{LE}). In order to achieve this, we place two conditions on the 
solutions of equation (\ref{LE}). First,  
let us impose that our bulk dynamics produces a surface perpendicular
to the $z$-axis without overhangs and bubbles. This condition is necessary
to ensure that we have a well defined interface (note that equation 
(\ref{LE}) contains overhang/vacancy and shadowing effects).
Secondly, we shall neglect any
interfacial profile. Then, we may assume that the solutions of the Langevin
equation (\ref{LE}) have the form
\begin{equation}
\rho({\bf r},t)=\Theta\big(h_t({\bf x})-z\big) \label{sol}
\label{profile}
\end{equation}
where ${\bf r}\equiv({\bf x},z)$,
${\bf x}\in R^{d}$ is a point in the substrate and
$h_t({\bf x})$ is the height of the growing surface
at time $t$.
$\Theta(\lambda)=1,1/2,0$ if 
$\lambda>0,=0,<0$ respectively.
That is, for a given point in the bulk $\bf{r}$, if its $z$ coordinate is 
larger, equal or smaller than the actual position of the surface, $h_t({\bf x})$,
then the density
field is $\rho({\bf r},t)=0$, $1/2$ and $1$ respectively.
Note that since the $\Theta$ function is not continuous, 
when differenciating
we should use a regularized version of it, e.g.
$\Theta(x)=\frac{1}{2} [1+
\tanh(ax)]$ with
$a \to \infty$. 

We are interested in constructing a dynamical equation for
the $h_t({\bf x})$ fields. Therefore, 
let us make a time derivative in equation (\ref{sol})
\begin{equation}
\partial_t\rho({\bf r},t)=\partial_th_t({\bf x})\delta(h_t({\bf x})-z).
\label{dersol}
\end{equation}
Integrating in $z$ both sides of equation (\ref{dersol})
we find
\begin{equation}
\partial_th_t({\bf x})=\int_R dz \ \partial_t\rho({\bf r},t). \label{p1}
\end{equation}
Equating this expression for $\partial_t h_t({\bf x})$, together with 
(\ref{LE}), will lead us to the desired Langevin equation for the heights.

To make this a bit more concrete, we now introduce a particular set of
rates. 
For instance, we choose the probability of
adding mass to the point $\bf r$ to be proportional to the 
square of the gradient of the density field in that point:
$w(\rho;{\bf r}) \propto |\nabla \rho |^2$.
With this election  
the unwanted effect of nucleation of bubbles is avoided.
After a bit of algebra, we get 
\begin{eqnarray}
\partial_t h_t=&&
\alpha \Big(1+(\nabla h_t)^2\Big)+
\frac{D}{2\Omega}\Delta h\nonumber \\
&& \qquad +\frac{1}{\Omega^{1/2}}
\Big(1+({\nabla} h_t)^2 \Big)^{1/2}\nu_t, 
\label{KPZ2}
\end{eqnarray}
which is the celebrated KPZ equation with a different noise term
(a  naive power counting argument reduces the relevant part of (\ref{KPZ2}) 
to the KPZ equation).
The coefficient $D$ has the proper dimensions
and $\alpha$ is positive and depends
on how the $\Theta$ function is regularized.
This comes from our particular ansatz,
but, we would like to stress here that, as far as universal properties
are concerned, the precise value of the coefficients is immaterial.
In fact, it is easy to show (with naive power counting)
that for any bulk dynamics
given by $w(\rho;r) \propto |\nabla \rho |^{\eta}$ with
$\eta \ge 0$, gives rise to a height equation falling in
the KPZ universality class. 

Next, we proceed to check numerically the connection between
eq.(\ref{MEGM}) with $w(\rho;{\bf r}) \propto |\nabla \rho |^2$ and
the KPZ equation (\ref{KPZ2}).


{\it Numerical results.}
The simple bulk rate $|\nabla \rho |^2$ can be easily implemented in a 
Monte Carlo experiment. On a two-dimensional square lattice 
periodic boundary conditions are considered in one of the principal axes.
Each lattice site is labeled by an occupation variable $\rho_{\bf r}$
ranging from $0,1/\Omega,...$ to 1. A site is empty if $\rho_{\bf r}=0$
and full if $\rho_{\bf r}=1$.
Initially the lattice is empty except for a full horizontal bottom line.
The growth starts when an 
empty site $\bf r$ is chosen at 
random from the lattice. Then, $\rho_{\bf r}$ is increased in $\Omega^{-1}$
with a probability that depends on each nearest neighbour of $\bf r$ 
through $|\nabla \rho_{\bf r}|^2$. 
Since symmetrized discrete forms of the gradient operator 
may give rise to  bulk vacancies incompatible with (\ref{profile}),
we use the following finite-difference
formula for the density derivatives 
$({\bf r }=(x,z))$
 
\begin{equation}
\nabla \rho_{\bf r} =\Big(\rho(x\pm1,z)-\rho(x,z),\rho(x,z)-\rho(x,z-1)\Big).
\end{equation}
That is,
left and right derivatives are used alternately to avoid asymmetric
effects and, for convenience,
a unit lattice spacing is assumed.
The height is defined
as the distance to the highest occupied lattice site
directly above the substrate coordinate $x$.

\begin{figure}
\epsfig{file=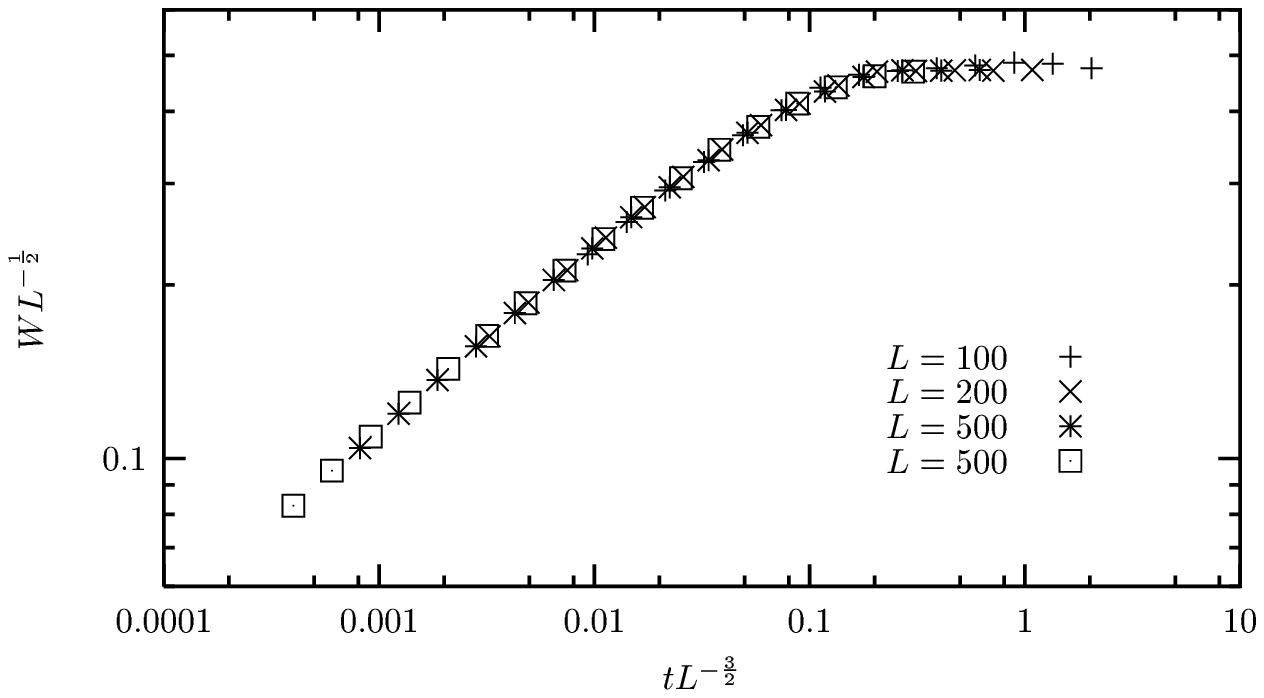,width=1.5\linewidth} \caption{Scaling plot
 for the model defined by the Master equation 
(\ref{MEGM}) with the rate $|\nabla \rho|^2$. 
The data ara for $L=$ 100, 200, 500 and 800, where 
$L$ is the width of ths substrate.}
\label{fig2}
\end{figure}

Figure \ref{fig2} shows the 
scaling plot for the surface width 
$W(L,t)^2=L^{-1} \sum[h_t(x)-\bar h (t)]^2$
for different system substrate sizes $L$.
$\bar h (t)$ is the mean height of the interface
at time $t$.
The numerical data were averaged over 2000 independent runs
for $L=100, 200, 500$ and over 1000 independent runs for $L=800$.
For reasons of computational efficiency the results shown correnspond 
to $\Omega=1$. Other values of $\Omega$
yield similar results but the simulations are much more time-consuming.
Good data colapse is obtained for a roughness exponent $\alpha=1/2$ and 
a dynamic exponent $z=3/2$, in agreement
with the KPZ prediction \cite{KPZ}.

The example we have just provided is by no means unique. 
Our formalism emcompasses
many others well known growth equations.
Let us just mention that with the simple dynamics given by
$w(\rho,{\bf r})=|\nabla \rho|$ the equation of 
 Golubovic and Wang, related to the anharmonic equilibrium thermal
fluctuations of smectics A \cite{GW}, is obtained. This is 
given by 
\begin{eqnarray}
\partial_t h_t=
 \Big(1+(\nabla h_t)^2\Big)(\lambda + \frac{1}{\Omega}  H) \\
  +\frac{1}{\Omega^{1/2}}
\Big(1+({\nabla} h_t)^2 \Big)^{1/4}\nu_t, 
\label{GWE}
\end{eqnarray}
where $\lambda $ is a coefficient and
$H$ is the curvature (see \cite{GW}).
As we have mentioned before, this kind of dynamics
(proportional to $|\nabla \rho|^{\eta}$ with $\eta \ge 0$),
falls in the KPZ universality class. Also,
the Edwards-Wilkinson equation \cite{ew}, which favours growth at
local minima, can also be
recovered by considering substraction of particles and a rate of
the form $w=|\nabla^2 \rho|$.
In this case, the formalism has to be slightly modified by 
linking the election of additon or substraction of particles to the 
density field configuration. More explicitely,
now the Master equation defining the process reads

\begin{eqnarray}
\partial_{\tau} P_{\tau}({\bf \rho})=&&\int_{R^{d+1}}d{\bf r}\biggl[
c({\bf \rho'}\rightarrow
{\bf \rho})P_{\tau}({\bf \rho}')\nonumber\\
&& \qquad \qquad -
c({\bf \rho}\rightarrow{\bf \rho}')P_{\tau}({\bf \rho})
\biggr],
\end{eqnarray}
with $\rho'=\rho({\bf x}) + \alpha \Omega^{-1} \delta({\bf x}
-{\bf r})$ and $\alpha= -1,0,1$ for $\nabla^2 \rho$ less, equal 
and greater than 0, respectively. That is, material is added to those areas
where the laplacian of the density field is negative and taken from
those where it is positive. In this manner, a balanced 
distribution of mass is achieved that results in the equilibrium 
Edwards-Wilkinson universality class.

Many other different rates lead to their corresponding growth 
equations, sometimes to the same one, showing that
growth models with similar surface behavior may not have the 
same bulk properties. 

\section{Conclusions}

In this paper we have introduced 
a class of nonequilibrium models in which a stochastic bulk
dynamics is defined. The bulk evolves
by an adsorption process represented by
a continuum Master equation. From it, we have derived
a Langevin equation
for the bulk density field
whose structure depends on the details of the 
underlying bulk dynamics. This dependence was then extended to an 
equation of motion for the interfacial degrees of freedom. In particular, 
we have examplified the procedure by deriving the  KPZ equation 
from a very simple bulk rate. A Monte Carlo simulation of
the bulk process confirms the predicted scaling behavior for the interface.
Finally, a number of examples were briefly mentioned. 
In all cases the bulk
dynamics determines the mesoscopic height equation, 
showing that both scales are related in 
a non trivial form and that their mutual influence could be far from intuitive.

The strategy developed in this paper is quite general. It includes
both local and nonlocal, and equilibrium and nonequilibrium growth
processes. Therefore, a great number of growth physical phenomena
can be studied, in principle, with our approach. 
For instance, molecular beam epitaxy
models with adatom mobility, driven lattice gases 
or wetting phenomena by means of lattice
gas theories of multilayer adsorption, to name just a few.

\acknowledgments

We acknowledge
E. Hern\'andez-Garc\'\i a
and  M.A. Mu{\~n}oz
for a critical reading of the manuscript.
C.L. acknowledges the hospitality received at the Centro de F\'\i sica da
Mat\'eria Condensada da Universidade de Lisboa and is
 supported by CICYT (MAR98-0840). F.S. is supported by the European 
Comission under grant ERBFMRXCT980183.

\end{multicols}

\begin{references}

\bibitem{G} J. Krug and H. Spohn  in 
{\it Solids far from equilibrium}, ed. C. Godreche,
Cambridge University Press,  1992.

\bibitem{KV} J. Kert{\'e}sz and T. Vicsek, {\it Self affine interfaces} in {\it Fractals
in Science} A. Bunde and S. Havlin eds., p. 89, Springer-Verlag, 1994.

\bibitem{BS} A.L. Barabasi and H.E. Stanley, {\it Fractal concepts in surface
growth}, Cambridge University Press, 1995.

\bibitem{HZ} T. Halpin-Healy and Y.C. Zhang,
Phys. Rep. {\bf 254}, 215 (1995).

\bibitem{LDS} Z.W. Lai and S. D. Sarma,
Phys. Rev. Lett. {\bf 66}, 2348 (1991).

\bibitem{MMTB} M. Marsili, A. Maritan, F. Toigo and J.R. Banavar,
Rev. Mod. Phys. {\bf 68}, 963 (1996).

\bibitem{KPZ} M. Kardar, G. Parisi and Y.C. Zhang,
Phys. Rev. Lett. {\bf 56}, 889 (1986).

\bibitem{vicsek} T. Vicsek, M. Cserz\"o and V.K. Horv\'ath,
Physica A {\bf 167}, 315 (1990).


\bibitem{KBR} R.P.U. Karunasiri, R. Bruinsma and J. Rudnick,
Phys. Rev.
Lett., {\bf 62}, 788  (1989).

\bibitem{DLGK} S. Das Sarma, C.J. Lanczycki, S.V. Ghaisas and
J.M. Kim, Phys. Rev. B {\bf 49}, 10693 (1994).

\bibitem{SK} M. Schimschak and J. Krug, Phys. Rev. B {\bf 52},
8550 (1995).


\bibitem{spohn} H. Spohn, J. Stat. Phys. {\bf 71}, 1081 (1993).


\bibitem{GM} P.L. Garrido and J. Marro, J. Phys. A {\bf 25}, 1453 (1992);
J. Stat. Phys. {\bf 74}, 663 (1994).   

\bibitem{HH} P.C. Hohenberg and B.I. Halperin, Rev. Mod. Phys. {\bf 49}, 435
(1977).

\bibitem{GSM} P.L. Garrido, M.A. Mu{\~n}oz and F. de los Santos, Phys. Rev.
E, {\bf 61}, R4683 (2000).

\bibitem{GGG} B. Grossmann, H. Guo and M. Grant, 
Phys. Rev. A, {\bf 43}, 1727 (1991).

\bibitem{KMTB} P. Keblinski, A. Maritan, F. Toigo and J.R. Banavar,
Phys. Rev. E {\bf 49}, R4795 (1994).

\bibitem{PC} G. Parisi and Y.C. Zhang,  
J. Stat. Phys. {\bf 41}, 1 (1985).



 
\bibitem{vkm}
N.G. van Kampen,
{\em Stochastic Processes in Physics and Chemistry\/},
North Holland, 1992.


\bibitem{KU} T.G. Kurtz, 
Stochastic Proc. and Appl. {\bf 6}, 223 (1978).

\bibitem{GW} L. Golubovic and Z.G. Wang,
Phys. Rev. Lett. {\bf 69}, 2535 (1992).


\bibitem{ew}
S.F. Edwards and D.R. Wilkinson, Proc. R. Soc. London, Ser. A, {\bf 381},
17 (1982).
\end{references}
\end{document}